\documentclass[prl,aps,twocolumn,]{revtex4}
\newcommand{\beq}{\begin{equation}}
\newcommand{\eeq}{\end{equation}}
\usepackage{graphics}
\usepackage{amsmath}
\usepackage{amsfonts}
\usepackage{amssymb}
\usepackage{graphicx}

\begin{document}

\title {Mechanism for the high Neel temperature in SrTcO$_3$}
\author{S. Middey$^1$, Ashis Kumar Nandy$^{1,2}$, Priya Mahadevan$^2$ and D.D. Sarma$^3$}
\affiliation
{$^1$Centre for Advanced Materials,Indian Association for the Cultivation of Science, Jadavpur, Kolkata-700032, India  \\
$^2$ S.N. Bose National Centre for Basic Sciences, JD-Block, Sector III, Salt Lake, Kolkata-700098, India. \\
$^3$ Solid State and Structural Chemistry Unit, Indian Institute of Science, Bangalore-560012, India \\
}
\begin{abstract}
The microscopic origin of the high Neel temperature ($T_N$) observed experimentally in SrTcO$_3$ has been examined using a 
combination of ab-initio electronic structure calculations and mean-field solutions of a multiband Hubbard model. 
The G-type antiferromagnetic state is found to be robust for a large region of parameter space, with large stabilization energies
found, surprisingly, for small values of intraatomic exchange interaction strength as well as large bandwidths. The microscopic
origin of this is traced to specific aspects associated with the $d^3$ configuration at the transition-metal site. Considering
values of interaction strengths appropriate for SrTcO$_3$ and the corresponding 3$d$ oxide SrMnO$_3$, we find a ratio of 4:1 for the
$T_N$ as well as magnitudes consistent with experiment.

\end{abstract}

PACS number(s): 75.10.-b, 75.47.Lx
\maketitle

The picture of magnetism that has prevailed over the years has centred around the existence of localized electrons and their ordering leading to different types of magnetic order. Consequently one associates the highest magnetic ordering temperatures with the more correlated 3$d$ transition metal oxides, with examples among the 4$d$ and 5$d$ 
oxides which have wider bands, more a rarity than the norm. It was therefore a surprise when recently a high magnetic ordering temperature ($T_N$) of 1023 K and 800 K were found  in  4$d$ transition metal oxides SrTcO$_3$~\cite{srtco3_prl} and CaTcO$_3$~\cite{catco3_jacs} respectively.  These $T_N$ were much higher than  any of their 3$d$ counterparts (SrMnO$_3$; $T_N$=233 K)~\cite{srmno3_camno3}. In a short span, these unexpected experimental observations have generated a lot of theoretical interest and several possible explanations were offered \cite{srtco3_prl,atco3_prb,srtco3_gorges}. One reason was the smaller Hund's coupling strength and the larger bandwidth associated with the 4$d$ oxides~\cite{srtco3_prl}. However, within a picture of itinerant magnetism that one has so far, both these correspond to effects which should result in a reduction in the magnetic ordering temperatures and so the puzzle remains. An alternate explanation offered by Georges and coworkers was that SrTcO$_3$ sits at the boundary between the itinerant to localized regime and hence has such a high transition temperature~\cite{srtco3_gorges}.

In this work we reexamine the issue of the high Neel temperature of SrTcO$_3$. An unusual aspect associated with the half-filling of the $t_{2g}$ levels results in a bandgap opening up for a small value of $U$ for the antiferromagnetic state, associated with the nesting of the fermi  surface. As a result, a magnetic moment is stabilized for small $U$ in the G-AFM state, while other magnetic solutions are able to sustain a magnetic moment only at larger values of $U$.
In the insulating state, there are channels present for the electrons to delocalize and lower their energy only in the 
antiferromagnetic state. This energy lowering which strongly stabilizes the G-AFM state is larger for small intra-atomic 
exchange interaction strength ($J_h$) as well as large hopping strength. 
The metal-insulator transition for a ferromagnetic state takes place at a larger value of $U$ than the G-AFM state. So, 
although there are channels for delocalization present in the ferromagnetic metallic state, these are smaller
than the G-AFM state and vanish as $U$ is increased and the system becomes insulating. As the energy gain by delocalization for the 
G-AFM state decreases from 5$d$ to 4$d$ to 3$d$ transition metal compounds, one expects a similar trend in the $T_N$. The 
large T$_N$ observed for SrTcO$_3$ and CaTcO$_3$ we show is a generic feature of all 4$d$ as well as 
5$d$ oxides with a formal $d^3$ configuration on the transition metal atoms.

The electronic and magnetic structure of  SrTcO$_3$ has been calculated within a plane wave pseudopotential implementation of density functional theory using PAW potentials~\cite{paw_potential} as implemented in VASP~\cite{vasp}. In addition to the GGA form for the exchange correlation functional, we also included an effective  $U$ of 2 and 3 eV  on Tc in the Dudarev implementation~\cite{dudarev_u_formalism} of the GGA + $U$ scheme as earlier work had shown that a value of 2.5 eV was appropriate for Ru~\cite{priya_prb_srruo3thinfilm}. The total energies were calculated for different magnetic configurations using a k-mesh of 6x6x6 k-points and a cut off energy of 400 eV for the plane wave basis states. In these calculations, the lattice constants were kept fixed at the experimental values~\cite{srtco3_prl}, while the internal coordinates were optimized to minimize the total energy. In order to 
understand the origin of the observed magnetic stability and its dependence on microscopic parameters, we carried out additional analysis in terms of a multiband Hubbard-like Hamiltonian. The parameters entering the tight binding part of the Hamiltonian are determined by  fitting the ab-initio band structure for the nonmagnetic case to a tight binding model that included $s$ and $p$ states on oxygen and $d$ states on Tc. Hopping is included between Tc $d$ and O $s$ as well as $p$ states, as well as between the $p$ states on oxygen atoms and these are parameterized in terms of the Slater-Koster parameters~\cite{slater_koster}. The semi-core O $s$ states were included to simulate the splitting 
between the Tc states with $t_{2g}$ and $e_g$ symmetry at $\Gamma$ point~\cite{mattheiss_prb}. The Coulomb matrix elements entering the multiband Hubbard-like part of the Hamiltonian are parametrised in terms of the Slater-Condon 
integrals $F^0$, $F^2$ and $F^4$~\cite{slater_condon}. $F^2$ and $F^4$ were scaled from their atomic values to result in a particular Hund's coupling strength $J_h$, while the value of $F^0$ was chosen to result in an effective Coulomb interaction strength equal to $U$. A value of $J_h$ = 0.1 eV has been used for most of the calculations in order to examine the small $J_h$ limit. The Hamiltonian was solved using a mean-field decoupling scheme for the four fermion terms till a convergence of 10$^{-5}$  was achieved on the energy~\cite{dd_hf}. The total energies determined for different magnetic configurations were mapped onto a Heisenberg model (-$\frac{1}{2}\sum J_{ij}$ s$_{i}$.s$_{j}$) with first neighbor ($J_1$) as well as second neighbor ($J_2$) exchange interaction strengths. An effective exchange interaction strength $J_0$ given by 6$J_1$+ 12$J_2$ was determined as this is directly related to $T_N$ in a mean field model, upto a multiplicative constant.

The total energies referenced to the nonmagnetic state for different magnetic configurations determined from our ab-initio calculations are given in Table I. As observed earlier~\cite{srtco3_prl}, the ferromagnetic and A-AFM calculations converge to a nonmagnetic solution for $U$ = 0. Surprisingly the magnetic structures in which 
each Tc atom has more number of antiferromagnetic neighbors are the ones which converge to a magnetic solution. 
This was also found earlier and a possible reason was attributed to a need for beyond LDA effects due to the incorrect treatment of the residual exchange-correlation effects in 4$d$ compounds~\cite{atco3_prb}. Correcting for 
this with HSE functionals, they were able to examine all magnetic solutions and discuss trends in the Neel temperature for all members of the series $A$TcO$_3$, where $A$ is an alkali metal atom. We use a computationally less intensive method, GGA + $U$ with a $U$ = 2 and 3 eV on Tc and are able to converge to magnetic solutions for all configurations. 
Stabilization energies comparable with the HSE results are obtained. Apart from the robustness of the G-AFM ground state, we also find a larger stability for the  G-AFM state over the FM state for $U$ = 2 than for $U$ = $3$. While it is indeed true that the antiferromagnetic stability due to superexchange processes is expected to reduce as $U$ is increased, it is surprising that both the magnitude of the stabilization energy as well as its variation with $U$ are substantial for small values of $U$.


\begin{table}
\caption
 {Stabilization energy (meV/f.u.) with respect to the nonmagnetic state}
\begin{tabular}{lcccc}
\hline
\hline
U(eV) & Ferro & A-AFM & C-AFM & G-AFM\\
\hline
0   & nonmag   & nonmag  & -46 & -168 \\
2  &-432 & -568 & -662 & -745 \\
3   & -925 & -1029 & -1100 & -1164\\
\hline
\end{tabular}
\end{table}

The reasons for observed trends in first principles calculations are usually difficult to pinpoint as a result of being dependent on several other parameters, which need not be the same as $U$ is varied. In order to carry out a microscopic analysis to understand the origin of magnetic ordering we set up a multiband Hubbard-like model for SrTcO$_3$
with a $U$ on Tc. The hopping matrix elements were parametrised in terms of the Slater Koster parameters $pd\sigma$, $pd\pi$, $sd\sigma$, $pp\sigma$ and $pp\pi$~\cite{slater_koster}. A least squared error minimization procedure was used to estimate the best set of parameters entering the tight-binding part of the Hamiltonian that best fit the ab-initio band structure~\cite{priya_tb}. The bands with primarily Tc $d$ character as well as the O $p$ nonbonding states were included in the fitting. The parameters that were obtained were $sd\sigma$ = -3.4 eV, $pd\sigma$  =-3.3 eV, $pd\pi$ = 1.55 eV, $pp\sigma$ = 0.6 eV and $pp\pi$ = -0.15 eV and $\epsilon_d$ - $\epsilon_p$ = 2 eV. 


\begin{figure}
\resizebox{7cm}{!}
{\includegraphics*[14pt,20pt][575pt,835pt]{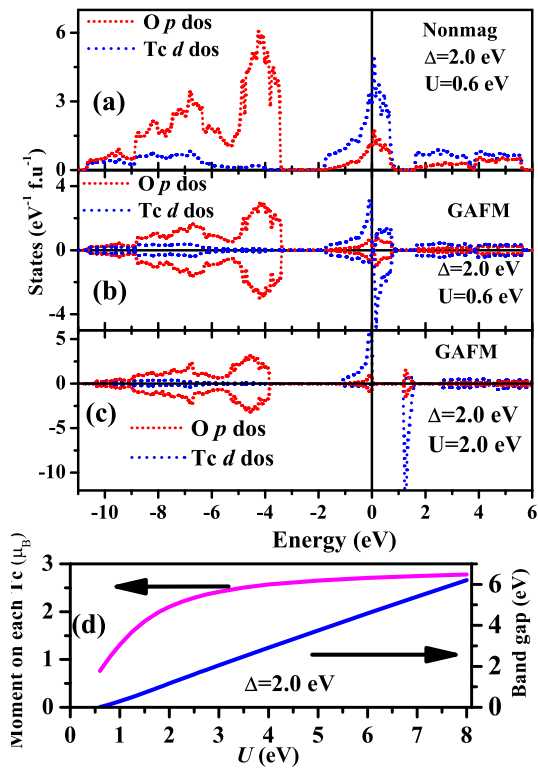}} \\
\caption{(color online) Partial density of states of  Tc  $d$ states and O $p$ states for (a) nonmagnetic solution with $\Delta$ = 2.0 eV, $U$ = 0.6 eV; (b)  GAFM spin configuration with $\Delta$ = 2.0 eV, $U$ = 0.6 eV; (c) GAFM spin configuration with $\Delta$ = 2.0 eV, $U$ = 2.0 eV. The variation band gap and magnitude of magnetic moment of each Tc atom with $U$ are plotted in (d) for $\Delta$ = 2.0 eV }
\end{figure}

After obtaining the hopping strengths, we solve the multiband Hubbard Hamiltonian for several values of $U$ and $\Delta$. Features of the solutions observed earlier in the context of the ab-initio calculations are observed here also. While the G-type antiferromagnetic solution is found to be robust, the existence of the other 
magnetic solutions depends on the value of $U$. Examining the different solutions at $\Delta$=2~eV and a small value of $U$ equal to 0.6 eV, we find that only the G-AFM solution exists while all other magnetic solutions converge to a nonmagnetic solution. Examining the Tc $d$ and O $p$ partial density of states (Fig. 1(a)), we find that the 
$t_{2g}$ states with primarily Tc $d$ character  are almost 3~eV wide. As a consequence of nesting associated with a bipartite lattice at  half-filling~\cite{fs_nesting}, a band gap opens up for just $U$ = 0.6 eV and $J_h$=0.1 eV for the G-AFM configuration (Fig. 1(b)). This increases to 1.1  eV when $U$ is increased to 2 eV (Fig. 1(c)), scaling almost linearly with $U$ as shown in Fig. 1(d). Once the system goes insulating, it can very easily sustain a local magnetic moment as the hopping processes between the majority spin $t_{2g}$ states at one site and the unoccupied minority spin $t_{2g}$ states at the neighboring site enhances the stability of the antiferromagnetic state. The variation of the magnetic moment in the G-AFM  configuration at each Tc site is plotted as a function of $U$ in Fig. 1(d). The magnetic moment is found to increase from a value of 0.6 $\mu_B$ at $U$ = 0.6 eV to an almost saturation value of 2.5 $\mu_B$ at $U$ = 3 eV, in contrast to a fully ionic value of 3 $\mu_B$. These values give a sense of
the itinerant nature of the magnetism. For the ferromagnetic case or the other antiferromagnetic configurations 
where some neighbouring spins are aligned ferromagnetically, the metal-insulator transition takes place at a larger value of $U$. This explains the robustness of the G-AFM solution in regions of the parameter space where the other magnetic solutions converge to nonmagnetic solutions. 

\begin{figure}
\resizebox{7cm}{!}
{\includegraphics*[0pt,35pt][590pt,840pt]{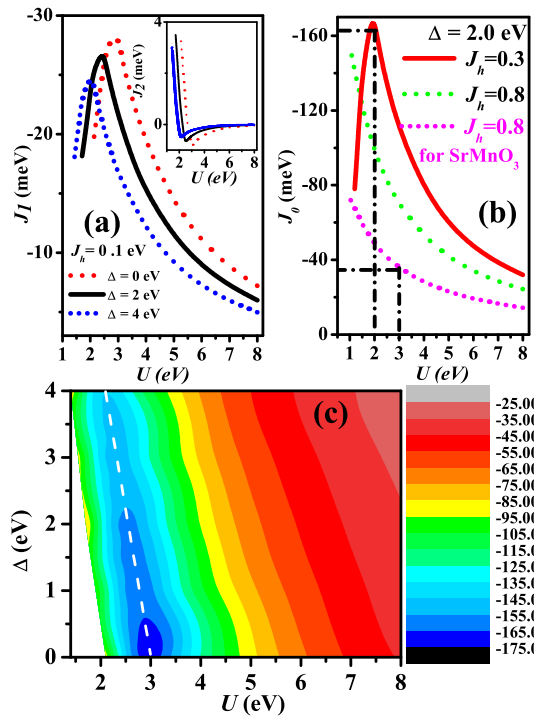}} \\
\caption{(color online) (a) Variation of $J_1$ with $U$ for $\Delta$ = 0, 2, 4 eV. The inset shows the corresponding variation of $J_2$. (b) $J_0$ as function of $U$ for $J_h$ = 0.3, 0.8 for SrTcO$_3$ and 0.8 eV for SrMnO$_3$. (c) Color plot of $J_0$  in $U$ - $\Delta$ plane for SrTcO$_3$ with $J_h$ = 0.1 eV. The dotted line indicates the position of maxima of $J_0$.}
\end{figure}

Two key parameters that control the properties of transition metal oxides are the charge transfer energy ($\Delta$) and the onsite Coulomb interaction strength ($U$). To check the stability of the magnetic state of SrTcO$_3$, we have calculated the interatomic exchange interaction strengths $J_i$'s with different combinations of $\Delta$ and $U$ with $J_h$ = 0.1 eV.  The variations in $J_1$ and $J_2$ as a function of $U$ for three  values of $\Delta$ equal to 0, 2 and 4 eV are shown in  Fig. 2(a). As the G-AFM is found to be the lowest energy solution for all values of $\Delta$, $J_1$ 
turns out to be antiferromagnetic while $J_2$ (inset of Fig. 2(a)) is primarily ferromagnetic. Considering the case of $\Delta$=4, we find that the magnitude of $J_1$ first increases, and then decreases as $U$ is increased. The magnitude of $J_2$ decreases over a small $U$ variation and may then be approximated to 0. As $J_2$ involves higher order hopping processes, its value is finite only when the hopping interaction strength is sizeable (i.e. the small $U$ regime). As $J_0$ is dominated by $J_1$, we discuss the variation in $J_1$ that we find in order to understand the dependence of $T_N$ on microscopic interaction strengths. We are not able to extract $J_i$'s in the region where only some of the configurations converge to magnetic solutions. A simple perturbative treatment of the energies 
was carried out considering a fully spin polarized ground state at each transition metal site and the allowed first excited state. This simple model is able to capture the basic physics of why $J_1$ goes through a maximum as a function of $U$. As there are no delocalization pathways for a ferromagnetic arrangement, the energy gain for the antiferromagnetic arrangement is directly related to $J_1$. In the low $U$ limit, the $U$ is treated as the perturbation to the hopping and this tells us that $J_1$ varies as $U$. In the large $U$ limit, the hopping interaction is treated as a perturbation and this results in $J_1$ varying as $1/U$. It is immediately clear that this would imply that $J_1$ should go through a maximum as a function of $U$ and this is indeed what is observed in Fig. 2(a) for various values of $\Delta$ studied, all of which show similar dependencies. The largest $J_1$ is seen for the smallest $\Delta$ as the effective hopping between the sites will be the largest there. In order to probe the role of $J_h$, we have calculated $J_0$ for two values of $J_h$ - a value of 0.3 eV~\cite{Jh_abinitio}, as well as a typical value of 0.8 eV for SrTcO$_3$ and SrMnO$_3$ and plotted them in Fig. 2(b). Although the $J_0$ variation shows the same trend for both values of $J_h$, there is a dramatic reduction in $J_0$ in going from $J_h$ of 0.3 eV to 0.8 eV. This stems from the fact that the effective exchange splitting approximately varies as $U+2J_h$. Similar trends are seen for SrMnO$_3$ for which we show the variation in $J_0$ only for $J_h$ = 0.8 eV. The parameters used for solving the multiband Hubbard Hamiltonian for SrMnO$_3$ were derived from a tight-binding fitting of the ab-initio band structure as described  for SrTcO$_3$. The main difference in the extracted parameters for the two systems is in the value of $pd\pi$. This is 25$\%$ smaller in SrMnO$_3$ compared to SrTcO$_3$. If a $U$ of 3~eV is believed to be appropriate, for SrMnO$_3$ and a slightly reduced value of 2 eV seems likely for SrTcO$_3$, the ratio of the calculated $J_0$'s and therefore $T_N$s are in the ratio 4:1, with that for SrTcO$_3$ being higher and consistent with experiment. Using the caculated $J_0$
for SrTcO$_3$ and assuming a mean-field scaling of 80$\%$, we get a $T_N$ of 1020 K, in very good agreement with experiment. The values of $J_0$ have been extracted in the complete $U$ - $\Delta$ plane, and plotted in Fig. 2(c). The largest values of $J_0$ are seen near $U$ = 3 eV for small $\Delta$ ($\sim$ 0.3 eV) and as a function of $U$ are found about the dashed line drawn with a slight slope to the line $U$ = 3. There are two dominant energetics which determine the magnitude of $J_0$. The first is the delocalization channels present in the antiferromagnetic arrangement and the second are the delocalization channels present in the ferromagnetic metallic state. As a result of nesting, while the antiferromagnetic solution goes insulating at small values of $U$, 0.6 eV for $\Delta$ = 2.0 eV, 
the ferromagnetic solution remains metallic upto a value of $U$ equal to 2.5 eV which is of the order of the bandwidth. Hence depending on the magnitude of the effective hopping parameter, one could be in a regime where the peak is close to the point where the ferromagnetic solution goes insulating or far from it.  With increasing $\Delta$ as the effective hopping strength between neighboring Tc site decreases, one finds a reduction  in J$_1$. Additionaly for a fixed value of $\Delta$, as $U$ is increased one has a maximum and then a decrease with all $J_0$'s merging to the same value at large $U$ when the hopping is too small to bring about any magnetic ordering.

\begin{figure}
\resizebox{7cm}{!}
{\includegraphics*[15pt,40pt][590pt,822pt]{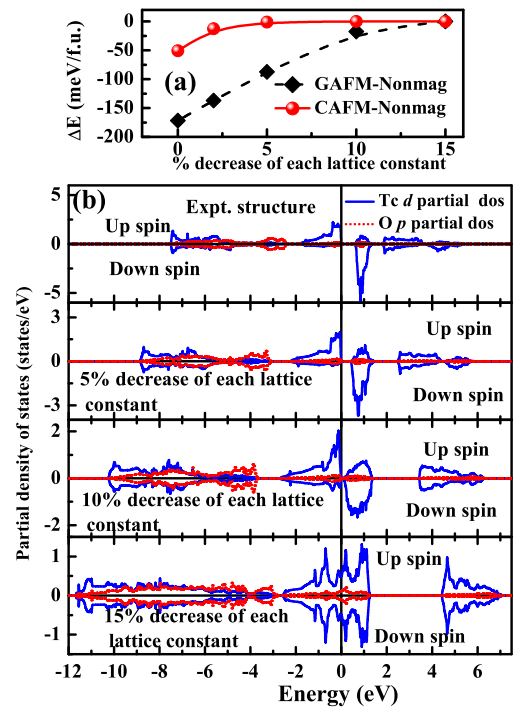}} \\
\caption{(color online) (a) Variation of magnetic stabilization energy for GAFM and CAFM spin confiurations with the percentage decrease of each lattice constant. (b) Tc $d$ and O $p$ partial density of states for GAFM configurations for different compressive strain. For 15\% decrease of lattice constant, the final spin configuration is nonmagnetic.}
\end{figure}

The results of Fig. 2 indicate a wide parameter regime for the stability of the G-AFM state as the ground state, 
thereby suggesting that the large magnetic stabilization energies are not limited to just SrTcO$_3$ and CaTcO$_3$, but
is generic of all transition metal oxides with a $d^3$ configuration. Another probe which could be used in experiments
is hydrostatic pressure which could be used to examine the stability of the G-AFM ground state. Evaluating the total
energy differences between the G- and the C-AFM  ground states within ab-initio GGA based calculations using the
SrTcO$_3$ structure as a starting point, we find that with a small change in the lattice constant of 2.5$\%$, the C-AFM
solution is no longer stable and converges to a nonmagnetic solution (Fig. 3(a)). The G-AFM solution we find is stable
and as revealed by the total energy difference with the nonmagnetic state plotted in Fig. 3(a) as well as the
Tc $d$ and O $p$ projected partial density of states plotted in Fig. 3 (b) for 0, 5, 10 and 14$\%$ decrease in
the lattice parameter. For a lattice parameter change larger than 10 $\%$ we find a collapse of the G-AFM solution to
the nonmagnetic solution, again supporting the robustness of the G-AFM solution.

We have examined the origin of a high $T_N$ in SrTcO$_3$ within the mean-field limit of a multiband Hubbard model. A wide range of parameters is found for which the G-AFM solution is stable. In contrast to usual expectations, the magnetic stabilization energy  is found to be larger in the limit of small $J_h$ and larger bandwidths. This is traced to aspects of the $d^3$ configuration at the transition metal site, which in the insulating state, allows electrons to delocalize only in the antiferromagnetic configuration. Using the appropriate values of the interaction strengths for SrTcO$_3$, as well as a mean-field reduction factor, we get a $T_N$ of 1020 K, which is four times larger than that for SrMnO$_3$.

SM, AKN thank CSIR, India for fellowship.







\begin{thebibliography}{99}
\bibitem{Jh_abinitio} From a fitting of the spin-polarized ab-initio band structure, $J_h$ for Tc is determined to be 0.3 eV.
\end{thebibliography}
\end{document}